      [1999/12/01 v1.4c Il Nuovo Cimento]
\documentclass{cimento}
\usepackage{graphicx}  
\title{ In-flight calibration of the SWIFT XRT}
\author{G.~Cusumano\from{ins:1}\ETC 
on behalf of the XRT Calibration Team.
}
\instlist{\inst{ins:1}INAF - Istituto di Astrofisica Spaziale e Fisica Cosmica di Palermo, Italy
}
\begin{document}

\maketitle

\begin{abstract}
The calibration of the Swift XRT effective area has been performed by analyzing cosmic sources observed during
the in-flight calibration phase and by using laboratory results and ray-tracing simulations as a starting
point. This work describes performance of the recent release of ancillary response files (ARF v8).  
\end{abstract}

XRT supports four different read-out modes to cover the dynamic range and rapid variability expected from GRBs
afterglows. The switch between modes, performed automatically, minimizes
pile-up and  optimizes the
collected information as the flux of the afterglow varies. 
In Imaging mode the XRT produces an integrated image (no X-ray event recognition takes place)  which,
for a typical GRB flux, is highly piled up. No spectroscopy is therefore possible,  but a very accurate position
and a good flux estimate can be obtained. 
The Photodiode mode (PD) is designed for very bright sources and allows
to observe GRB with high timing
resolution (0.14 ms). However, this operational mode went lost because
of a damaged in the CCD caused by a
micro-meteorite.
The Windowed Timing (WT) mode (1--600 mCrab) is obtained by binning and compressing 10 rows into a single row, and
then reading out only the central 200 columns of the CCD. It covers the central 8 arcminutes of the
field of view and provides one dimensional imaging and full spectral capability with a time resolution of 1.8 ms.
The Photon Counting (PC) mode ($<$ 1 mCrab) allows full spatial and spectral resolution with a timing resolution 
of 2.5 seconds.
Each read-out mode has a dedicated ARF file that contains the mirror effective area, the filter
transmission, as well as the vignetting function and the Point Spread Function (PSF) correction (which depends on
the source location and of the size of the extraction region) and residual correction of the CCD quantum
efficiency.

In WT mode the Crab nebula was used as main calibrator.  A moderate pile-up due the high count rate was partially
mitigated extracting the off-pulse spectrum (mainly due to the nebula) with a phase-resolved selection. Figure 1
(left panel)
shows the residuals obtained fitting the Crab nebula with an absorbed power law plus absorption features
(Kirsch et al. 2005). The fit was carried out in the 0.45-10 keV energy range  
with the low energy boundary limited by the unacceptable increase in
the residuals present in the spectrum below 0.45 keV.
Such an increase is due to
redistribution matrix problems. 
In the 0.45-10 keV energy range the $\chi^2_{red}$ is 1.83 (620 d.o.f.)
and the mean systematic uncertainty was at a level of 3\% (this is the
systematic uncertainty to be added in order to obtain in the Crab
spectrum a $\chi^2_{red}$ =1.0).
The strongest features visible in the Crab spectrum are mainly
below 2 keV.  The main systematics is a broad absorption features
at $\sim$0.5 keV. Residuals are also present at $\sim$1.5 keV and at 
$\sim$1.8 keV, while the systematics above 0.9 keV is due
to the pile-up effect. 
There is also some instrumental Nickel contamination
present in the 7-8 keV energy range which (sometimes) is not fully
subtracted. The edge-like residuals could be
due to energy scale offsets. We recommend to fit XRT WT spectra in the
0.3-10 keV energy range.

ARFs relative to the PC mode  were mainly calibrated with the SNR B0540-69. The strongest features in the SNR 0540-69 spectrum (see
Figure 1, right panel) are an
absorption feature at $\sim$0.5 keV and an emission feature at slightly higher energies. The latter feature is
large and might be related to the source itself as XMM-Newton spectra might indicate. Smaller features at larger
energies are also present. The statistical uncertainty on the final RMF+ARF matrices in PC mode is estimated at 3\%
level in the 0.3-10 keV energy range. 

\begin{figure}[t]
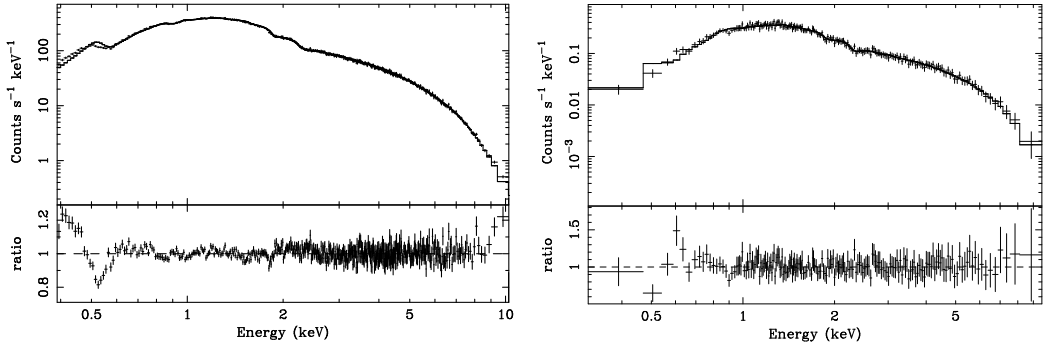

\hspace{-0.5truecm}
\includegraphics[width=4.5cm,origin=c,angle=-90]{cusumanog_fig1.ps} 
\hspace{0.1truecm} 
\includegraphics[width=4.5cm,origin=c,angle=-90]{cusumanog_fig2.ps}
\vspace{-1.1truecm}
\caption{{\bf Left panel:} 
The off-pulse phase resolved Crab spectrum (WT mode, grades 0-2), best fit model and the data/model ratio. 
{\bf Central panel:} 
The spectrum of SNR B0540-69  (PC mode, grades 0-12), best fit model and the data/model ratio.
\label{fig:1}}
\end{figure}

The on-going in-flight calibration has allowed to improve the effective area files for all observing modes and
grade selections to a level that satisfies the mission requirements.  
Our current knowledge of the XRT response still implies a systematic
uncertainty of the order of 3\% in the 0.3-10
keV energy band and of about 10\% in absolute flux. The following considerations
apply to both WT and PC mode observations.
For highly absorbed sources the response model showed an
under-estimation of the redistribution below about 1 keV.  
This effect
is clearly evident for $N_H>$ 10$^{22}$ cm$^{-2}$ but even for
less absorbed sources small deviations are present.
In case of bright sources we do experience small energy scale problem at 
low energies ( $E<$1 keV). This
problem is still under investigation and might be related to bright Earth contamination in PC mode and a bias
subtraction problem in WT mode causing energy scale offsets.

\end{document}